\documentclass{jfm}
\usepackage{graphicx}
\usepackage{amsmath}
\usepackage{bm}
\usepackage{epstopdf, epsfig}

\shorttitle{Gyrotactic swimmers in turbulence: shape effects and role of the large scale flow}
\shortauthor{Borgnino, Boffetta, De Lillo and Cencini}

\title{Gyrotactic swimmers in turbulence: shape effects and role of the large scale flow}

\author{M. Borgnino\aff{1}
  \corresp{\email{matteo.borgnino@unito.it}},
  G. Boffetta\aff{1},
  F. De Lillo\aff{1}
 \and M. Cencini\aff{2}\corresp{\email{massimo.cencini@cnr.it}}}
\affiliation{
\aff{1}Dipartimento di Fisica and INFN, Universit\`a di Torino, via Pietro Giuria 1, 10125 Torino, Italy
\aff{2} Istituto  dei  Sistemi  Complessi,  CNR,  via  dei  Taurini  19,  00185  Rome,  Italy  and  INFN  ``Tor  Vergata''
}

\begin{document}

\maketitle

\begin{abstract}
We study the dynamics and the statistics of dilute suspensions of
gyrotactic swimmers, a model for many aquatic motile microorganisms.
By means of extensive numerical simulations of the Navier-Stokes
equations at different Reynolds numbers, we investigate preferential
sampling and small scale clustering as a function of the swimming
(stability and speed) and shape parameters, considering in particular
the limits of spherical and rod-like particles.  While
spherical swimmers preferentially sample  local downwelling
flow, for elongated swimmers we observe a transition from
downwelling to upwelling regions at sufficiently high swimming speed.
The spatial distribution of both spherical and elongated swimmers is
found to be fractal at small scales in a wide range of swimming
parameters.  The direct comparison between the different shapes shows
that spherical swimmers are more clusterized at small stability and
speed numbers, while for large values of the parameters elongated
cells concentrate more. The relevance of our results for phytoplankton
swimming in the ocean is briefly discussed.
\end{abstract}

\begin{keywords}
%Authors should not enter keywords on the manuscript, as these must be chosen by the author during the online submission process and will then be added during the typesetting process (see http://journals.cambridge.org/data/\linebreak[3]relatedlink/jfm-\linebreak[3]keywords.pdf for the full list)
\end{keywords}

%%%%%%%%%%%%%%%%%%%%%%%%%%%%%%%%%%%%%%%%%%%%%%%%%%%%%%%%%
\section{Introduction}
%%%%%%%%%%%%%%%%%%%%%%%%%%%%%%%%%%%%%%%%%%%%%%%%%%%%%%%%

Understanding the dynamical and statistical properties of
self-propelled agents (such as motile microorganisms, microrobots or
phoretic particles) in a flow is key to a range of fields encompassing
aquatic ecology \citep{kiorboe2008mechanistic,guasto2012fluid},
biomedicine \citep{nelson2010microrobots} and active matter
\citep{marchetti2013hydrodynamics}. Recently, microswimmers have
gathered much attention in virtue of their collective behavior
\citep{marchetti2013hydrodynamics,elgeti2015physics}. However, in the
presence of a background flow, as  typical in
natural and artificial aquatic environments, the dynamics of
microswimmers can be highly non-trivial even at the level of the
single swimmer or for dilute (non interacting) suspensions. Indeed
complex behavior can originate from the interplay between swimming,
advection, the fluid velocity gradients that act on the particles by
changing their direction of propulsion and the possible presence of
biases (such as chemotaxis, gravi/gyro-taxis, phototaxis, etc.)
affecting the swimming direction \citep[for a review see,
  e.g.,][]{guasto2012fluid}. For instance,
\citet{Torney2007,Torney2008} were among the first to put forward the
possibility of fractal clustering and non-trivial mixing properties
for elongated microswimmers, with or without phototaxis, in a
Taylor-Green vortical flows. \citet{rusconi2014bacterial} found that
cell elongation is responsible for aggregation and transport
suppression of bacteria in shear flows due to Jeffery orbits
\citep{jeffery1922motion}. Reduced transport and vortical trapping was
observed for spherical microswimmers in a chaotic flow
\citep{khurana2011reduced}.

In this work we focus on gyrotactic swimmers in turbulent flows.  Many
motile phytoplankton species are gyrotactic, i.e. they swim in the
direction resulting from the competition between stabilizing torque
due to buoyancy/gravity and the shear-induced viscous torque
\citep{kessler1985hydrodynamic,Pedley1992}. The stabilizing torque can
originate either from bottom heaviness or from a fore-rear asymmetry
\citep{ten2014gravitaxis,sengupta2017phytoplankton}, which tend to
keep the swimming direction oriented upwards favoring vertical
migration toward well-lit waters near the surface.  Like other forms
of biased motility gyrotaxis can impinge the transport properties and
spatial distribution of swimming plankton.  In both laminar and
turbulent pipe flows gyrotaxis leads to peculiar transport properties
with respect to passive particles
\citep{bearon2012biased,croze2013dispersion}. In laminar flows, it
induces remarkable beam-like accumulations in downwelling pipe flows
\citep{kessler1985hydrodynamic} and in vortical flows
\citep{cencini2016centripetal}, and high-concentration layers in
horizontal shear flows \citep{durham2009disruption,Santamaria2014}.
In moderately turbulent flows, as in the oceans, gyrotaxis can
generate intense microscale fractal patchiness
\citep{durham2013turbulence,zhan2014accumulation,fouxon2015phytoplankton,gustavsson2016preferential} and, in strong turbulence, accumulation in vortical regions
\citep{delillo2014turbulent}. Moreover, gyrotactic cells in turbulence
have been found to preferentially sample specific flow regions
\citep{durham2013turbulence,fouxon2015phytoplankton,gustavsson2016preferential}.
Indeed, direct numerical simulations (DNS) in both turbulent
(multiscale) and stochastic (single scale) flows, corroborated by
theoretical considerations, have shown that spherical gyrotactic cells
preferentially visit downwelling regions of the flow
\citep{durham2013turbulence,fouxon2015phytoplankton,gustavsson2016preferential},
which can hinder the upward vertical cell migration.
\citet{gustavsson2016preferential} also found a dependence on cell's
morphology in stochastic flows: elongated swimming particles can
preferentially sample either downwelling or upwelling flow regions,
depending on their swimming speed and stability. This is rather
intriguing since some gyrotactic cells can change their morphology
and biasing direction when exposed to turbulence
\citep{sengupta2017phytoplankton} and in this way control their
motility. Moreover, \citet{zhan2014accumulation} and
\citet{gustavsson2016preferential} found that turbulent suspensions of
elongated gyrotactic cells are generally less clustered than spherical
ones but for the regime in which gyrotaxis is very weak.

Here we aim at clarifying the differences in the dynamics of turbulent
suspensions of spherical and elongated gyrotactic cells by means of
DNS.  The rest of the paper is organized as follows.  In
Sec.~\ref{sec:model}, we present the model and the numerical
procedure. In Sec. \ref{sec:pref} and \ref{sec:fractal} we present our
results on preferential sampling of the vertical component of fluid
velocity and on fractal clustering, respectively. We end with
Sec.~\ref{sec:conclusions} by summarizing and discussing our findings.

%%%%%%%%%%%%%%%%%%%%%%%%%%%%%%%%%%%%%%%%%%%%%%%%%%%%%
\section{Model equations and previous results}\label{sec:model}
%%%%%%%%%%%%%%%%%%%%%%%%%%%%%%%%%%%%%%%%%%%%%%%%%%%

Following \citet{kessler1985hydrodynamic} and
\citet{Pedley1992}, we model gyrotactic swimmers as axisymmetric
ellipsoids swimming with constant speed $v_s$ in the direction ${\bf
  p}$ ($|{\bf p}|=1$), along the axis of symmetry of the
ellipsoid. Owing to their size being typically smaller than the
Kolmogorov scale and their density being very close to that of the
carrier fluid, the microswimmers behave as tracers if not for the
motility. Neglecting interactions between swimmers (as
appropriate for dilute suspensions) and stochastic reorientations, the
evolution of the position, $\bm x$, and orientation, ${\bf p}$, of
each swimmers is ruled by the following equations:
\begin{eqnarray}
\dot{\bm x}&=&{\bm u}({\bm x},t)+\Phi {\bf p} \label{eq:1}\\
\dot{\bf p}&=& {1 \over 2 \Psi} \left[\hat{\bf z}-(\hat{\bf z} \cdot {\bf p}) \bf p
\right] + {1 \over 2} {\bm \omega} \times {\bf p} + \alpha
\left[\hat{S}\mathrm{\bf p}-(\mathrm{\bf p}\cdot \hat{S}\mathrm{\bf
    p})\mathrm{\bf p}\right],
\label{eq:jeff}
\end{eqnarray}
where ${\bm u}$, ${\bm \omega}=\nabla\times{\bm u}$ and $\hat{S}_{ij}=
\frac12 (\partial_ju_i+\partial_iu_j)$ are the fluid velocity,
vorticity and the rate of strain tensor at the swimmer's position.
The three terms on the r.h.s. of (\ref{eq:jeff}) are: the
buoyancy/gravity torque tending to rotate the cell upwards (along the
vertical unit vector $\hat{\bf z}$) with a characteristic time $B$;
rotation due to vorticity and strain rate. The latter, so-called
Jeffery term \citep{jeffery1922motion}, is proportional to
$\alpha=(l^2-d^2)/(l^2+d^2)$ (with $l$ the length of the cell, along
${\bf p}$, and $d$ its width) and is absent for spherical cells
($\alpha=0$). In general $-1<\alpha<1$, with the extremes
corresponding to flat disks and rods, respectively.  In what follows,
we will consider only $\alpha\geq 0$, thus restricting ourselves to
the case of spherical or prolate cells, being the typical shapes of
microorganisms.   We remark that in (\ref{eq:jeff})
  we have not included a rotational stochastic term
  \citep{hill1997biased} due to the biological reorientation of the
  swimming direction \citep{Polin2009} since this effect is expected
 to be negligible with respect the reorientation induced by the small-scale
  turbulent flow.

As for the fluid velocity field, $\bm u$, we will consider moderately
turbulent flows described by the Navier-Stokes equation for
incompressible fluids ($\bm \nabla\cdot \bm u=0$)
\begin{equation}
\label{eq:ns}
\frac{\p {\bm u}}{\p t}+{\bm u}\cdot\nabla{\bm u}=-\nabla p+
\nabla^2{\bm u}+{\bm f},
\end{equation}
where $\nu$ is the kinematic viscosity, $p$ the pressure rescaled 
by the fluid density. 
The turbulent flow is maintained in a
statistically stationary state by the volume force ${\bm f}$, which
injects energy at rate $\epsilon$, statistically equal to the rate
of dissipation.  Based on the flow properties one can define the usual
small-scale parameters, namely the Kolmogorov scale
$\eta=(\nu^3/\epsilon)^{1/4}$, time $\tau_\eta=(\nu/\epsilon)^{1/2}$
and velocity $u_\eta=(\nu\epsilon)^{1/4}$.  The large dynamics can be
parameterized in terms of the rms value of the velocity $u_{\rm
  rms}=\sqrt{\langle|{\bf u}|^2\rangle/3}$ (with $\langle...\rangle$
denoting ensemble average).
In the following physical variables will be made dimensionless
by rescaling spatial and temporal coordinates with the Kolmogorov scale
$\eta$ and time $\tau_{\eta}$ respectively, and velocities with the
Kolmogorov velocity $u_{\eta}$. 
Correspondingly, swimmer motion will be governed by the dimensionless
swimming number $\Phi=v_s/u_\eta$ and stability number $\Psi=B/\tau_\eta$,
while the flow is parameterized in terms of the Taylor-scale Reynolds number
${\rm Re}_\lambda=u_{\rm rms}\lambda/\nu$ where
$\lambda=\sqrt{15\nu/\epsilon}u_{\rm rms}$.

Based on the model equations (\ref{eq:1},\ref{eq:jeff}), recent
numerical and theoretical works have provided insights into the
properties of clustering of gyrotactic swimmers in both turbulent and
chaotic flows \citep{durham2013turbulence,zhan2014accumulation,
  fouxon2015phytoplankton,gustavsson2016preferential}. In particular,
it was first shown in \cite{durham2013turbulence} that spherical
gyrotactic swimmers in turbulence can form fractal clusters. A
phenomenological argument for this goes as follows.  In the limit of
fast orientation $\Psi\to 0$, cells tend to swim upwards (${\bf p}\to
\hat{\bf z}$) so that their trajectories would approximately follow
the effective velocity field ${\bf v}={\bm u}+\Phi\hat{\bf z}$, which
is incompressible, $\bm \nabla\cdot{\bf v}=0$, and thus they do not
cluster. Also in the limit of strongly unstable cells, $\Psi\gg 1$,
there would be no accumulation, since the swimming direction would be
essentially random.  However, when $\Psi\ll 1$ (and $\Phi\ll 1$) a
perturbative solution of (\ref{eq:jeff}) suggests (in analogy to
inertial particles in the limit of small Stokes number
\citep{balkovsky2001intermittent,Bec2003}) that the effective velocity
field, ${\bf v}={\bm u}+\Phi{\bf p}^*$ with ${\bf p}^*=(\Psi \omega_y,
-\Psi\omega_x, 1)$, is compressible with $\bm \nabla\cdot {\bf
  v}=-\Psi\Phi\nabla^2u_z$.  Hence, for $\Psi\ll 1$ cells behave as
tracers in a compressible velocity field, and evolve onto a fractal
attractor.  On the basis of this argument, one concludes that cells
will preferentially sample downwelling portions of the flow, where
$u_z<0$ \citep{durham2013turbulence} \citep[see][for a refined
  derivation]{fouxon2015phytoplankton}. The same conclusion was
obtained with a perturbative approach in a stochastic single scale
velocity field in \cite{gustavsson2016preferential}. In the latter
paper, however, the authors consider the general case of elongated
swimmers, $\alpha\geq 0$. Analytical results and simulations of
stochastic model indicate that while preferential sampling of
downwards velocities is typical of the limit $\max(\Phi,\Psi)\ll 1$
for small aspect ratio swimmers, this is reversed for
$\alpha>\alpha_c=3/5$, i.e. sufficiently elongated swimmers
concentrate in upwelling regions.  Furthermore, the inversion is
predicted to be present only for $\Phi>\Phi_c(\Psi)$, with a critical
swimming parameter decreasing with $\Psi$.

%%%%%%%%%%%%%%%%%%%%%%%%%%%%%%%%%%%%%%%%%%%%%%%%%%%%%%%%%
\subsection{Numerical simulations} \label{sec:numerics}
%%%%%%%%%%%%%%%%%%%%%%%%%%%%%%%%%%%%%%%%%%%%%%%%%%%%%%%%%
The Navier-Stokes equations (\ref{eq:ns}) are solved in a three
dimensional, periodic cubic domain of size $L=2 \pi$ 
by means of a parallel, fully dealiased pseudo-spectral code
at resolutions of $64^3$, $256^3$ and $1024^3$ grid
points, corresponding to  ${\rm Re}_\lambda=21,68,173$.
A constant energy input is provided by a 
deterministic, large scale forcing ${\bm f}$ acting on modes with
wavenumbers in the spherical shell $1<|{\bf k}|<3$.
Time integration is implemented by a second order Runge-Kutta
scheme with exact integration of the dissipative term. 
Accuracy at small scales is guaranteed
by ensuring that $k_{max}\eta\gtrsim1.9$ for all simulations.

In each run, the trajectories of up to $128000$ swimmers per parameter
set are integrated according to (\ref{eq:1},\ref{eq:jeff}). The
Eulerian velocity is interpolated at the
particle positions by third order polynomials. The
  derivatives needed for the calculation of ${\bm \omega}$ and
$\hat{S}_{ij}$ are computed by using the derivatives of the
interpolating polynomials thus ensuring that they are second-order
accurate.  For each ${\rm Re}_\lambda$ we have explored a wide range
of cells' parameters, with $\Psi \in \left[0.1 : 50 \right]$, $\Phi
\in \left[0.1 : 30 \right]$ and shape parameter $\alpha=0,0.5,1$ thus
considering up to 90 different types of swimmers per simulation.
Swimmers are initialized with uniform random positions $\bm x$ in the
domain and orientations $\bm p$ on the unit sphere.  At stationarity,
data are collected for several configurations ($120-200$) separated by
about half a large-scale eddy turnover time to ensure statistical convergence.

%%%%%%%%%%%%%%%%%%%%%%%%%%%%%%%%%%%%%%%%%%%%%%%%%%%%%%%%%
\section{Results}
%%%%%%%%%%%%%%%%%%%%%%%%%%%%%%%%%%%%%%%%%%%%%%%%%%%%%%%%%

%%%%%%%%%%%%%%%%%%%%%%%%%%%%%%%%%%%%%%%%%%%%%%%%%%%%%%%%%
\subsection{Preferential sampling of fluid velocities} \label{sec:pref}
%%%%%%%%%%%%%%%%%%%%%%%%%%%%%%%%%%%%%%%%%%%%%%%%%%%%%%%%%

\begin{figure}
  \centerline{
\includegraphics[width=1\textwidth]{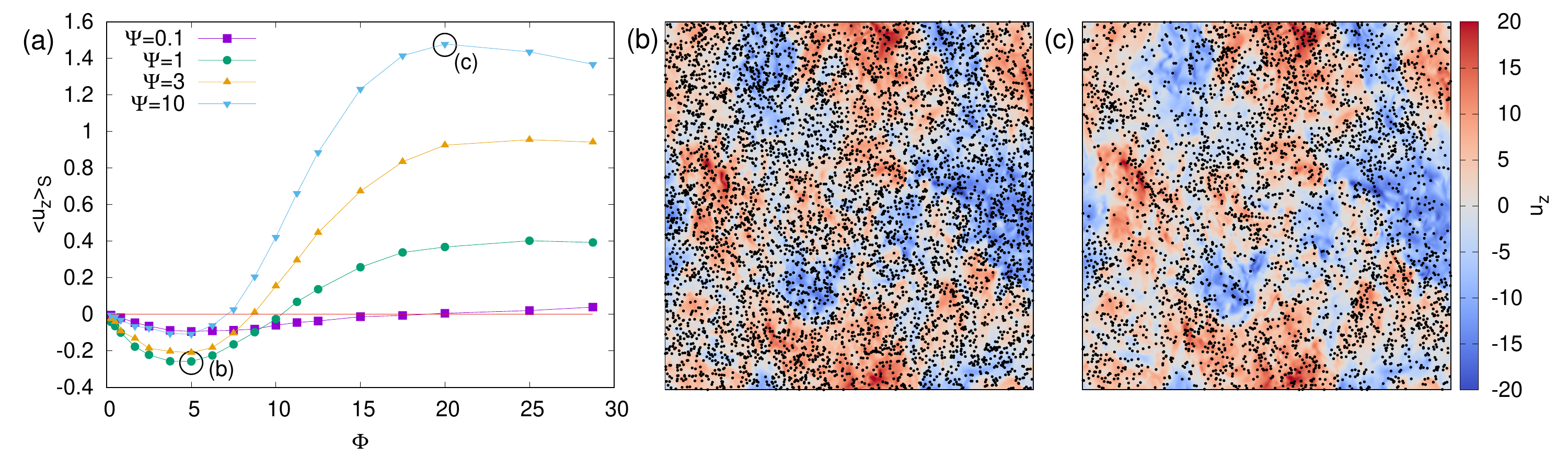}}
  \caption{Preferential sampling of the vertical component of fluid velocity by
gyrotactic, rod-like swimmers ($\alpha=1$). (a) Average vertical component of
the fluid velocity measured along cell trajectories. Curves are computed at
${\rm Re}_\lambda=173$. Panels (b) and (c): distributions of swimmers 
for populations marked in panel (a) with $\Psi=1$,
$\Phi=5$ (b) and $\Psi=10$, $\Phi=20$ (c) plotted on a color map
of the vertical velocity $u_z$. Slower (b) or faster (c) swimmers 
sample preferentially regions with negative (blue) or positive (red) vertical 
velocity.}
\label{fig:uz_phi}
\end{figure}

\begin{figure}
\centerline{
\includegraphics[width=0.9\textwidth]{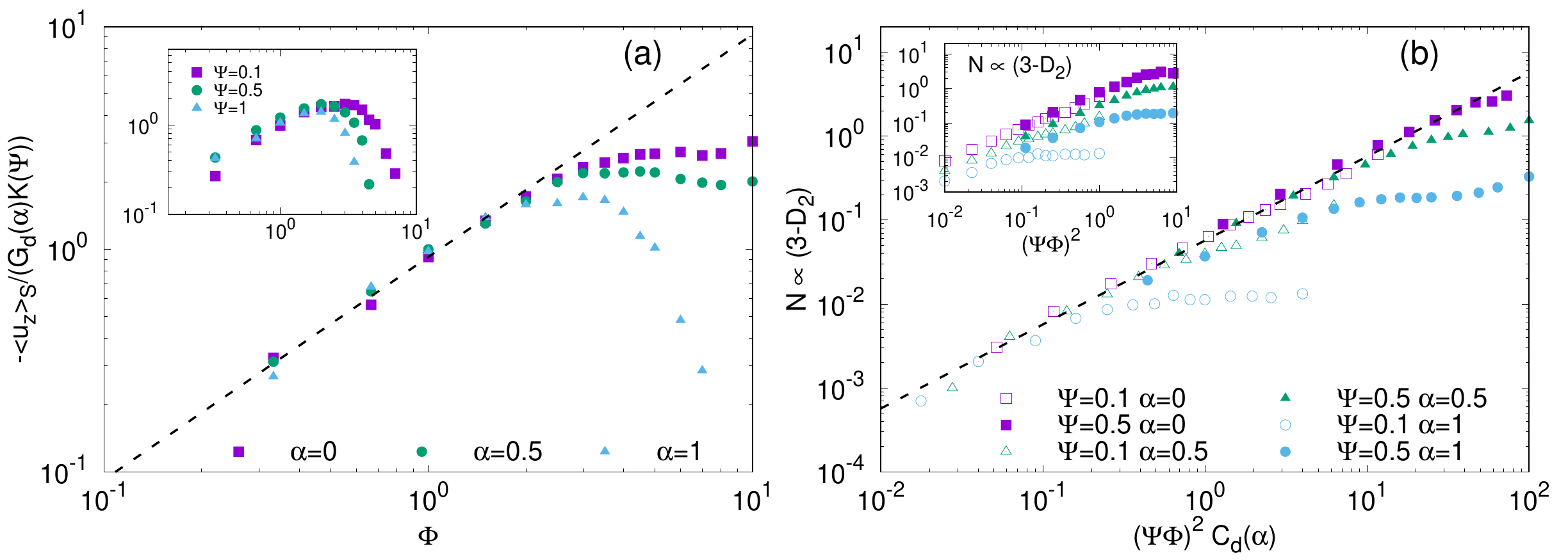}
}
\caption{Small $\Phi$ behavior of preferential sampling of fluid
  velocities and small scale clustering, compared with the theoretical
  predictions in \cite{gustavsson2016preferential} (dashed lines). a)
  vertical component of fluid velocity conditioned on swimmer
  trajectories for different elongations as labeled
  at $\Psi=0.1$. Inset: same as main panel but for rods ($\alpha=1$)
  at different $\Psi$.  b) Accumulation index $N$ rescaled with the
  stochastic prediction. Inset: the same curves without rescaling.}
\label{fig:mehligphi}
\end{figure}

\begin{figure}
  \centerline{ \includegraphics[width=1.0\textwidth]{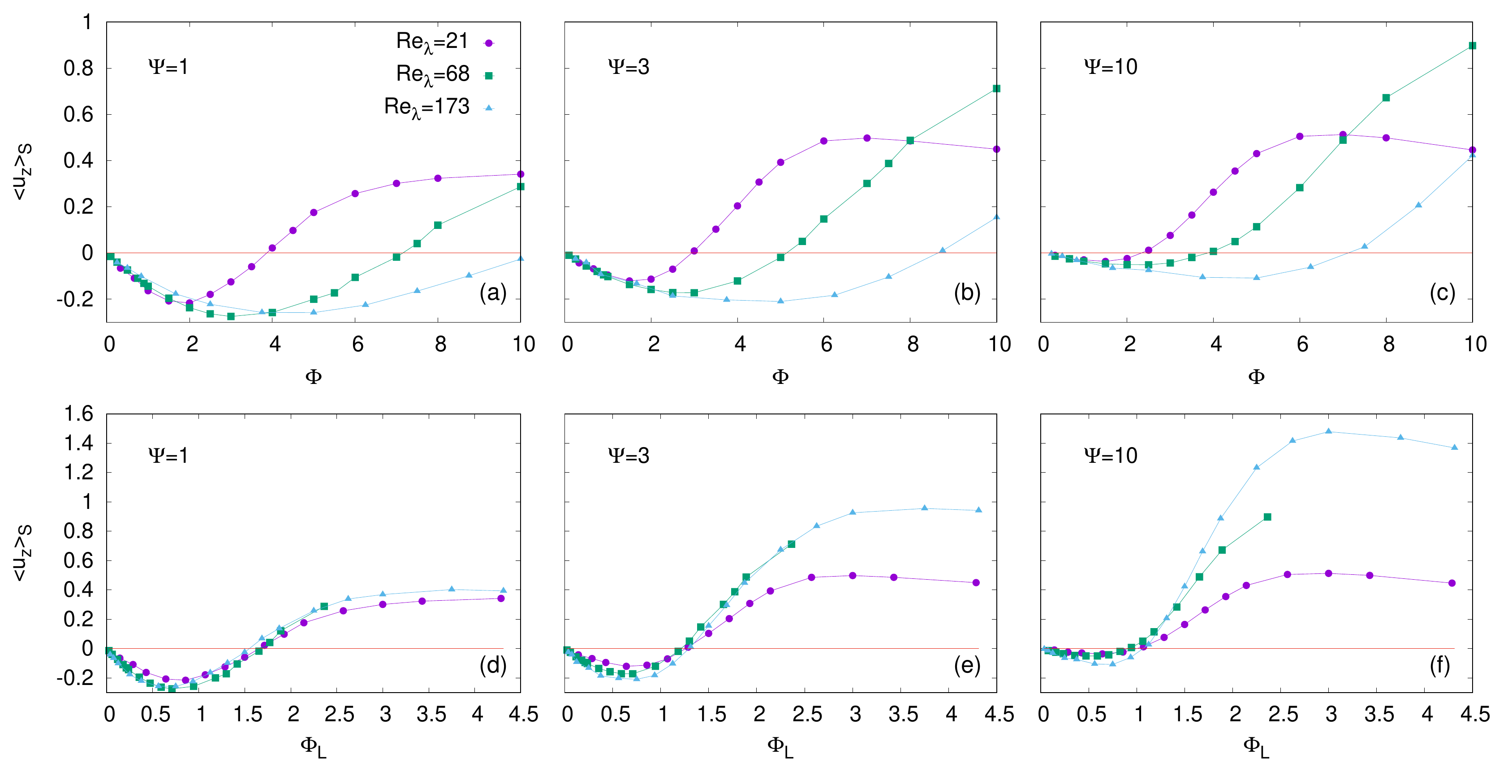}}
  \caption{Average of the vertical fluid velocity, $\langle
    u_z\rangle_S$, along swimmer trajectories for
      rod-like ($\alpha=1$) cells as a function of $\Phi=v_s/u_\eta$
    (panels (a), (b) and (c)), and of $\Phi_L=v_s/u_{\rm rms}$ (panels
    (d), (e) and (f)) for different ${\rm Re}_\lambda$, as labeled in
    (a). While the small $\Phi$ behavior in the top panels rescales
    consistently, the transition to preferential sampling of positive
    values of $u_z$ is better rescaled by $\Phi_L$.}
\label{fig:uz_psi_Re}
\end{figure}

The vertical displacement in the water column of a gyrotactic swimmer
following (\ref{eq:1}) can be affected by fluid transport even in a
flow with vanishing average velocity.  Indeed, although the Eulerian
average of the fluid velocity vanishes, the average swimmer velocity
$\langle {\bm v}\rangle=\langle {\bm u} \rangle_S
  +\Phi\langle \mathrm{p}_z\rangle \hat{\bf z}$ depends on the mean
value of ${\bm u}$ computed along particle trajectories
(notice that in the above expression symmetry implies
  $\langle \mathrm{p}_{x,y}\rangle=0$).  While for the
horizontal components one has $\langle u_x\rangle_S=\langle
u_y\rangle_S=0$ for symmetry reasons, gravitaxis breaks the up-down
symmetry so that one can have $\langle u_z\rangle_S\ne 0$
\citep{fouxon2015phytoplankton}. Since the largest swimming speeds of
motile algae are typically of the order of $u_\eta$ or smaller and
$u_{\rm rms}\gg u_\eta$, preferential sampling of regions of upwards
or downwards flow can in principle strongly enhance or hinder the
swimmer's drift towards the surface.  Figure \ref{fig:uz_phi}(a) shows
the average value of the vertical component of the fluid velocity
along swimmer trajectories $\langle u_z\rangle_S$. The different
curves are computed for rod-like cells ($\alpha=1$) and $\langle
u_z\rangle_S$ is plotted for several values of $\Psi$ as a function of
the swimming parameter, $\Phi$. It is clear that slow swimmers tend to
preferentially sample downwards fluid velocities
(figure~\ref{fig:uz_phi}(b)), while faster ones
(figure~\ref{fig:uz_phi}(c)) spend more time where $u_z>0$.

For a more quantitative analysis, we focus at first on small values of
$\Phi$. In this limit the behavior of the curves is consistent with
what observed in \citep{durham2013turbulence}, in that the average
fluid velocity seen by the cells is increasingly negative as cell
velocity grows. This behavior has been discussed for a
$d$-dimensional, random, single-scale flow in
\citep{gustavsson2016preferential} where it was predicted that in the
small $\Phi$ limit, $\langle u_z\rangle_S
\propto-\Phi G_d(\alpha)K(\Psi)$, where
$G_d(\alpha)=[d(1-\alpha)+2]/d$ and $K(\Psi)=\Psi/(2\Psi+1)$.  That
prediction is remarkably verified in our simulations, as shown in
figure \ref{fig:mehligphi} where the average fluid velocity along the
trajectories is plotted for different values of the shape parameter
$\alpha$, showing that the factor due to particle geometry
$G_d(\alpha)$ is indeed correct. When the shape parameter $\alpha$ is
fixed, the dependence on $\Psi$ is also fairly well satisfied (see
inset). A similar prediction is confirmed also on the small $\Phi$
properties of clustering as shown in figure 2(b), which will be
discussed in the next section.  In general, we can also conclude that
the small $\Phi$ behavior of this observable is independent of ${\rm
  Re}_\lambda$. Panels (a-c) of figure \ref{fig:uz_psi_Re} show the
vertical fluid speed curves for fixed $\Psi$ at various ${\rm
  Re}_\lambda$. Notice that in our nondimensional
  units, based on the Kolmogorov velocity, the curves collapse in the
  small $\Phi$ region for each value of $\Psi$ \citep[see also][for
  the case of spherical cells]{durham2013turbulence}.

\begin{figure}
  \centerline{
\includegraphics[width=.60\textwidth]{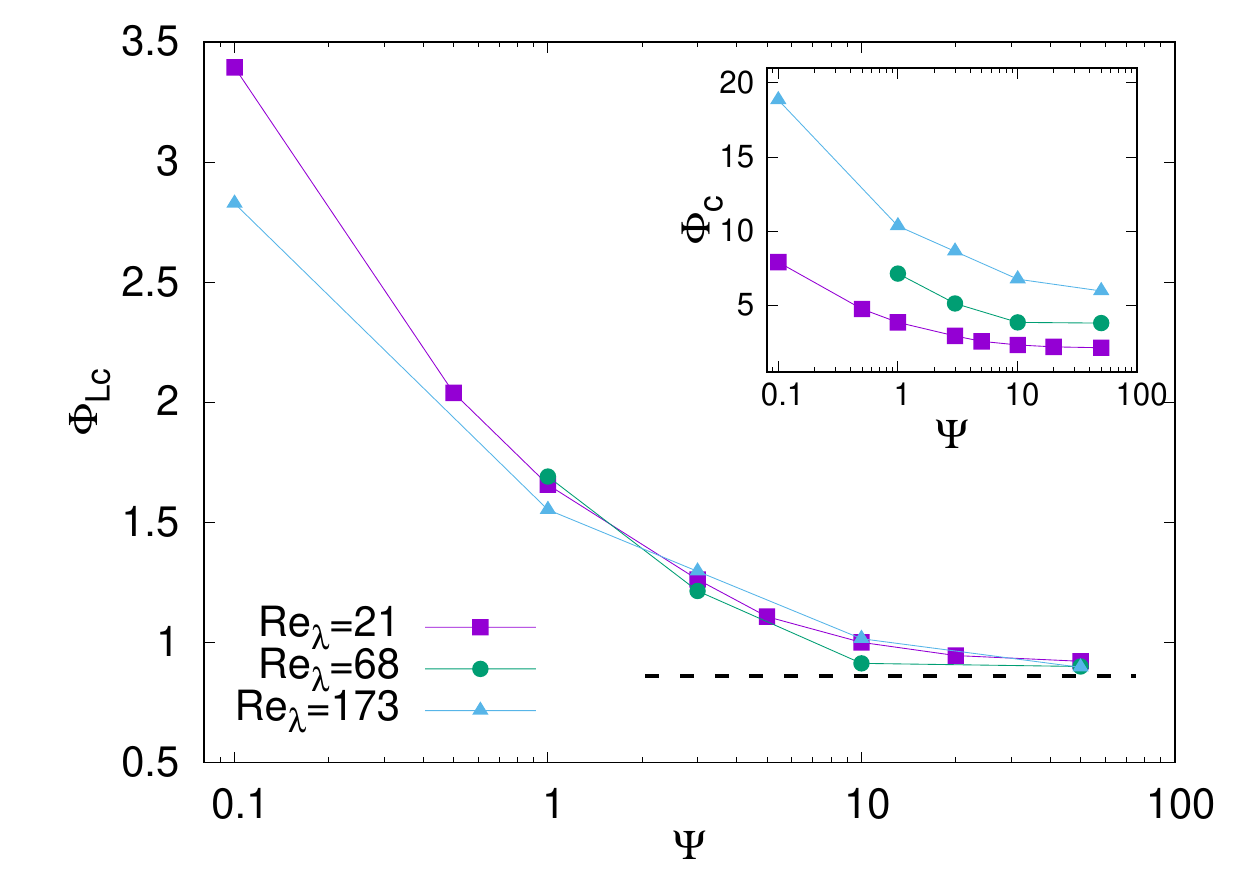}}
  \caption{Critical value of $\Phi_L$ (main panel)
      and $\Phi$ (inset) for various ${\rm Re}_\lambda$ as a function
      of $\Psi$. Dashed line: theoretical asymptotic value of
    $\Phi_L,c$ (see text).}
\label{fig:phiCpsi}
\end{figure}

We now consider larger values of the swimming velocity.  As noted
above, fast swimming cells tend to preferentially sample regions of
upwards flow. As shown by Fig.~\ref{fig:uz_psi_Re} the critical speed
$\Phi_c$ of transition from downwelling to upwelling regions (defined
as the point at which $\langle u_z \rangle_S=0$) is smaller for larger
values of $\Psi$.  However numerical results show that $\Phi_c$
saturates to a finite value for large $\Psi$. This has profound
implications for the statistics of velocity and velocity gradients
along particle trajectories.  While the correlation time of gradients
is of the order of $\tau_\eta$, the time it takes for a swimmer to
cross the correlation length of gradients $\eta$ is of the order of
$\tau_\Phi=\eta/v_s=\tau_\eta/\Phi$, so that $\Phi>1$ implies that the
gradients correlation time as seen by a swimmer is $\tau_{\rm
  corr}=\min(\tau_\Phi,\tau_\eta)=\tau_\Phi$
\citep{fouxon2015phytoplankton}. As a consequence, the preferential
sampling of upwards velocities cannot be understood in terms of a
quasi equilibrium solution, at variance with the $\langle
u_z\rangle_S<0$ case for spherical particles. However, an even more
important difference is revealed if the actual observed values of
$\Phi_c$ are considered. Indeed, when $\langle u_z\rangle_S$ is
plotted at fixed $\Psi$ upon changing the extension of the inertial range with simulations at different resolutions, one
can see that $\Phi_c$ increases with ${\rm Re}_\lambda$, and the
corresponding values of the swimming velocity are of the order of (or
larger than) the large-scale velocity of the flow $u_{\rm
  rms}$. Panels (d-f) of figure \ref{fig:uz_psi_Re} show the $\langle
u_z\rangle_S$ as a function of a large-scale swimming parameter
$\Phi_L=v_s/u_{\rm rms}$.  In this parameterization, the critical
values $\Phi_{L,c}(\Psi)$ are independent of ${\rm Re}_\lambda$. For
large $\Psi$, the critical $\Phi_L$ appears to asymptotically converge
from above to $\Phi_{L,\infty}\approx 0.86$, which can be obtained as
the $\Psi\gg 1$ limit of the analytical expression in
\citet{gustavsson2016preferential}. This numerical result is
summarized in Fig.~\ref{fig:phiCpsi} and is consistent with a picture
in which the transition to upwards velocity sampling is essentially
controlled by the integral scale of the flow.

The fact that the inversion in preferential sampling is controlled by
the large scale velocity was not noted before, to our knowledge.
Indeed it can be observed only if a sufficient separation between
large and small scales is present in the velocity field.  It is
therefore important to note that this scale separation effect is not
in contradiction with the stochastic results in
\citep{gustavsson2016preferential}, and indeed the asymptotic value of
the critical swimming parameter defined with respect to
$u_\mathrm{\rm rms}$ is in quantitative agreement with the prediction
found in the same paper (dashed line in figure~\ref{fig:phiCpsi}).
The techniques adopted in that paper relied on a flow with a single
scale, so that both gradients and velocities have the same correlation
times. It is remarkable that those techniques are able to capture a
qualitative behavior that is observed in a multiscale flow. The
observation that those features, in high ${\rm Re}_\lambda$ flows, are
found for very large swimming velocities (of the order of $u_{\rm
  rms}$ instead of $u_\eta$) can perhaps suggest why the qualitative
agreement between stochastic theory and simulations is so
good. Indeed, for very fast swimmers, as observed above, fluid fields
can be described as noise with correlation time inversely proportional
to the swimming speed, losing the details of the multiscale structure
of the flow.

%%%%%%%%%%%%%%%%%%%%%%%%%%%%%%%%%%%%%%%%%%%%%%%%%%%%%%%%%
\subsection{Fractal clustering} \label{sec:fractal}
%%%%%%%%%%%%%%%%%%%%%%%%%%%%%%%%%%%%%%%%%%%%%%%%%%%%%%%%%

We now focus on the small scale, fractal clustering properties of
gyrotactic suspensions at varying the shape parameter, $\alpha$, with
emphasis on the spherical ($\alpha=0$) and rod-like ($\alpha=1$)
limits. We start considering the limit of stable ($\Psi\ll 1$) and
slowly swimming ($\Phi\ll 1$) cells.

In such limit, the perturbative approach originally developed in
\citet{durham2013turbulence} for spherical cells (see
Sec.~\ref{sec:model}) can be extended to generic shapes by imposing
stationarity in (\ref{eq:jeff}) \citep{gustavsson2016preferential}.
At the first order in $\Psi$ one obtains that gyrotactic cells behave
as tracers in an effective velocity field ${\bm v}$
\begin{equation}
\dot{\bm x}={\bm v}={\bm u}+\phi {\bf p}^*\,, \quad \mathrm{with} \quad {\bf p}^*=[
\Psi(\omega_y+2\alpha\hat{S}_{13}),\Psi(-\omega_x+2\alpha\hat{S}_{23}),1]\,.
\end{equation}
The resulting velocity field, ${\bm v}$, is compressible with divergence
\begin{equation}
\nabla\cdot{\bm v}=-\Phi\Psi\left[\left(1+\alpha\right)\partial^2_zu_z+\left(1-\alpha\right)\left(\partial^2_x+\partial^2_y\right)u_z\right].
\end{equation}
For rods, this expression reduces to $\nabla\cdot{\bm
  v}=-2\Phi\Psi\partial^2_zu_z$, while for spheres it gives
$\nabla\cdot{\bm v}=-\Phi\Psi\nabla^2u_z$.
Tracers advected by a weakly compressible flow field ${\bm v}({\bm
  x},t)$ form clusters of fractal co-dimension $3-D\sim\langle
\nabla\cdot{\bm v}\rangle_S$
\citep{falkovich2002acceleration}. Thus one can infer that  swimmers should form
cluster with codimension $3-D\sim(\Phi\Psi)^2$. For a generic $\alpha$,
in a $d$-dimensional stochastic Gaussian flow $\langle \nabla\cdot{\bm
  v}\rangle_S$ can be computed exactly at first order of a
perturbative approach based on a Kubo expansion
\citep{gustavsson2016preferential}, obtaining the prediction 
\begin{equation}
3-D\sim(\Phi\Psi)^2C_d(\alpha)\qquad [C_d(\alpha)=[(d+2)(d+4)-2d(d+4)\alpha+(4+2d+d^2)\alpha^2]/d]\,.
\label{eq:predD2}
\end{equation}

We measured the fractal dimension of the clusters in terms of the
correlation dimension, $D_2$, ruling the small scale scaling behavior
of the probability to find a pair of cells with distance less than
$r$, i.e. $p_2(r)\sim r^{D_2}$ when $r\to 0$. We estimated $D_2$ from the local slopes of such probability.  When
$D_2\approx 3$, as expected in the limit $\Psi,\Phi\ll 1$ here
considered, this kind of measure is affected by large errors, so we
also measured the accumulation index $N$ which is obtained as follows.
The volume is divided in cubes of side $\ell$, and we count the
average number of particles in such cubes, $\langle n\rangle_\ell$,
and the standard deviation, $\sigma^2=\langle n^2\rangle_\ell-\langle
n\rangle_\ell^2$. Then we define $N=(\sigma -\sigma_P)/\langle
n\rangle_\ell$, where $\sigma_P=\sqrt{\langle n\rangle_\ell}$ is the
standard devision of a Poissonian process having the same mean number
of particles. As discussed in \citet{durham2013turbulence}, one can show
that if $\ell$ is small enough (here we have chosen $\ell \approx 2
\eta$) $N\propto 3-D_2$.

In figure~\ref{fig:mehligphi}b we show $N$ as a function of
$(\Psi\Phi)^2 C_d(\alpha)$ for different values of $\Psi,\Phi$ and
$\alpha=0,1/2,1$. We observe a fair collapse of the data (compare with
the inset where the geometric constant $C_d(\alpha)$ is removed) onto
a curve that displays a linear behavior for small values of the
abscissa, as predicted by Eq.~(\ref{eq:predD2}), providing again (see
also the discussion of figure~\ref{fig:mehligphi}a) strong evidence
that the first order Kubo expansion of
\citet{gustavsson2016preferential} reproduces the small $\Psi$ and
$\Phi$ limits of gyrotactic cells also in turbulent multiscale flows.
We observe that the linear limiting behavior seems to be more
persistent for the spherical shapes.  Similarly to inertial particles
which display the maximal clustering for Stokes number of order $1$
\citep{Bec2003}, gyrotactic cells also have a more pronounced
clustering for $\Psi\sim 1$
\citep{durham2013turbulence}. \citet{zhan2014accumulation} showed that
spherical cells are generally more clustered than elongated swimmers,
except for very large $\Psi$.  In Fig.~\ref{fig:D2} we compare
\textit{vis a vis} the behavior of $D_2$ (estimated from the local
slopes of $p_2(r)$) for spherical ($\alpha=0$) and rod-like
($\alpha=1$) as a function of the swimming number $\Phi$ for different
values of $\Psi\geq 1$. By comparing Fig.~\ref{fig:D2}a and c, it is
clear that spherical cells cluster more than elongated ones when
$\Psi$ is small enough, while the opposite is true for unstable cells
$\Psi\gg 1$. The latter limiting behavior can be explained by noticing
that at $\Psi\to \infty$ the dynamics of
Eq.~(\ref{eq:1}-{\ref{eq:jeff}) in the ($\bm x,{\bf p}$)-phase space
  has an average volume contraction rate equal to $-\alpha(d+2)
  \langle {\bf p}\cdot\hat{S} {\bf p}\rangle_S$, which is identically
  zero for spherical cells. In other terms, the dynamics of spherical
  cells with large $\Psi$ preserves volumes in phase space and so does
  its projection in position space (i.e. $D_2=3$) while elongated
  swimmers obey to a dissipative dynamics which lead to fractal
  clustering.  For intermediate values of $\Psi$
  (figure~\ref{fig:D2}b) a nontrivial role is played by the swimming
  speed: slow spherical swimmers cluster more than elongated ones, but
  the opposite is true for fast swimmers.

\begin{figure}
  \centerline{
\includegraphics[width=1.0\textwidth]{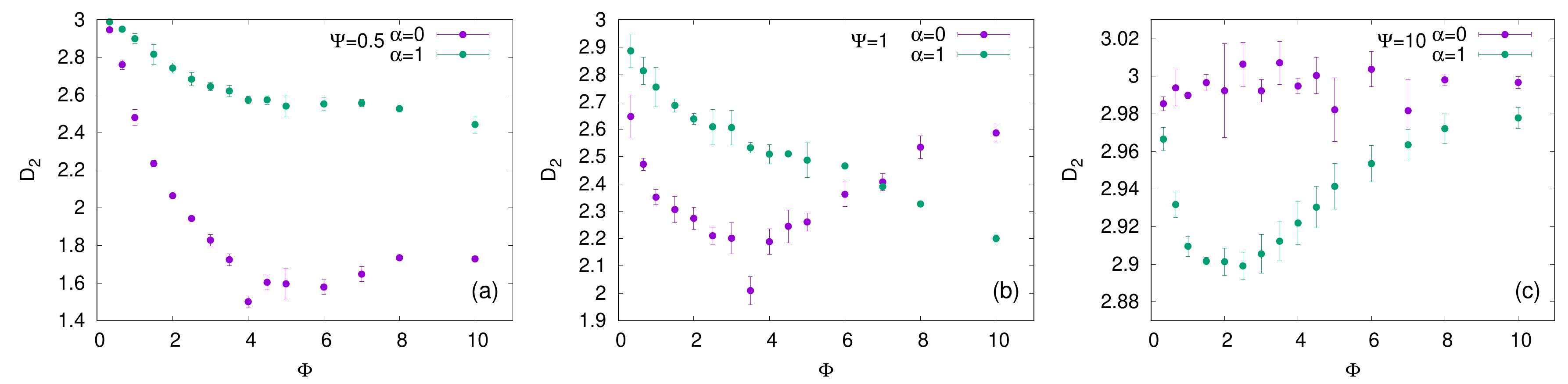}}
  \caption{Correlation dimension of clusters for $\Psi=0.5$ (a), $1.0$
    (b) and $10$ (c) as a function of the swimming parameter
    $\Phi$. Spherical swimmers ($\alpha=0$) concentrate more than
    elongated cells (in this case, ideal rods) for small values of
    $\Psi$ (a), while rod-like swimmers maintain observable clustering
    even for large $\Psi$ (c). For intermediate values of $\Psi$, slow
    (fast) elongated swimmers cluster less (more) than spherical
    ones (b). }
\label{fig:D2}
\end{figure}

%%%%%%%%%%%%%%%%%%%%%%%%%%%%%%%%%%%%%%%%%%%%%%%%%%%%%%%%%
\section{Conclusions and discussions\label{sec:conclusions}}
%%%%%%%%%%%%%%%%%%%%%%%%%%%%%%%%%%%%%%%%%%%%%%%%%%%%%%%%%
A systematic study of the statistical properties of gyrotactic
swimmers, at varying their shape (from spherical to rod-like) and
other cell's parameters, in realistic turbulent flows has been
presented. In particular, we focused on the characterization of their
aggregation in clusters and preferential sampling of the vertical
component of the fluid velocity field.  The first property is relevant
to many biological processes such as reproduction and predation
\citep{kiorboe2008mechanistic}, while the second impact on the
vertical migration of phytoplankton cells.  The main conclusions are:
(i) sufficiently slow swimmers (with swimming speeds $v_s \lesssim
u_\eta$), independently of their shape, always sample downwelling
region, the effect being maximal for non-dimensional reorientation
times, $\Psi$, of order $1$ and increasing with the swimming speed;
(ii) Fast elongated swimmers can sample upwelling regions (thus
acquiring an extra vertical migration speed), while this behavior is
absent for spherical cells (as previously shown in
  \citet{durham2013turbulence,gustavsson2016preferential} and
  confirmed here by simulations at large $\Phi$, not
  shown). Simulations at different Reynolds numbers show that the
transition from downwelling to upwelling regions is controlled by the
large scale velocity and thus requires $v_s \gtrsim u_{\rm rms}$;
(iii) Both spherical and elongated cells form fractal aggregates,
which are more clustered for spherical (elongated) cells in the regime
of fast (slow) reorientation. For intermediate reorientation, the
relative strength of clustering depends on the swimming speed.  (iv)
We provided evidence that some analytical results derived by
\citet{gustavsson2016preferential} for a stochastic flow are
qualitatively in agreement with numerical data obtained from fully
resolved turbulent flows. The agreement becomes quantitative in the
regime of slow swimming (Fig.~\ref{fig:mehligphi}).  This is
remarkable since the analytical results were obtained by a first order
Kubo-expansion for a single scale flow.

Our results clarify several aspects of the dynamics and the statistics
of gyrotactic suspensions and are of particular importance for
potential application to swimming phytoplankton, also in virtue of
recent evidence of possible morphological adaptations
\citep{sengupta2017phytoplankton}.  Most gyrotactic microorganisms
have reorientation times in the range $2-10 \, s$, with swimming speed
around $100-300\mu m/s$ and cell size around $10\mu m$
\citep{guasto2010oscillatory,harvey2015persistent}.  Typical values
for the energy dissipation rate in the ocean are in the range
$\epsilon \sim 10^{-4}-10^{-8}$ \citep{Thorpe2007} and therefore
phytoplankton is able to explore the parameter range $0.1 \lesssim
\Psi \lesssim 20-40$ and $0.1\lesssim \Phi\lesssim 3-5$ in which small
scale clustering is expected for both spherical and elongated
cells. On the contrary, given that the large scale velocity is
typically of order $cm/s$ or $m/s$ it is unlikely that phytoplankton
cells are able to reach sufficiently high $\Phi_L$ in order to exploit
preferential concentration in upwelling regions.  This regime could be
in principle accessible to faster natural or artificial swimmers, as
long as they are smaller than the Kolmogorov scale as required for the
validity of (\ref{eq:1}).

M.B., G.B. and F.D.L. acknowledge support by the {\it Departments of
  Excellence} grant (MIUR).  HPC center CINECA is gratefully
acknowledged for computing resources, within the INFN-Cineca agreement
INF18-fldturb.

%\bibliographystyle{jfm}
%\bibliography{biblio}

\begin{thebibliography}{32}
\expandafter\ifx\csname natexlab\endcsname\relax\def\natexlab#1{#1}\fi
\def\au#1{#1} \def\ed#1{#1} \def\yr#1{#1}\def\at#1{#1}\def\jt#1{\textit{#1}}
  \def\bt#1{#1}\def\bvol#1{\textbf{#1}} \def\vol#1{#1} \def\pg#1{#1}
  \def\publ#1{#1}\def\arxiv#1{#1}\def\org#1{#1}\def\st#1{\textit{#1}}

\bibitem[Balkovsky {\em et~al.\/}(2001)Balkovsky, Falkovich \&
  Fouxon]{balkovsky2001intermittent}
{\sc \au{Balkovsky, E.}, \au{Falkovich, G.} \& \au{Fouxon, A.}} \yr{2001}
  \at{Intermittent distribution of inertial particles in turbulent flows}.
  \jt{Phys. Rev. Lett.}  \bvol{86},  \pg{2790--2793}.

\bibitem[Bearon {\em et~al.\/}(2012)Bearon, Bees \& Croze]{bearon2012biased}
{\sc \au{Bearon, R.~N.}, \au{Bees, M.~A.} \& \au{Croze, O.~A.}} \yr{2012}
  \at{Biased swimming cells do not disperse in pipes as tracers: a population
  model based on microscale behaviour}.  \jt{Phys. Fluids}  \bvol{24}~(12),
  \pg{121902}.

\bibitem[Bec(2003)]{Bec2003}
{\sc \au{Bec, J.}} \yr{2003}  \at{Fractal clustering of inertial particles in
  random flows}.  \jt{Phys. Fluids}  \bvol{15},  \pg{L81}.

\bibitem[Cencini {\em et~al.\/}(2016)Cencini, Franchino, Santamaria \&
  Boffetta]{cencini2016centripetal}
{\sc \au{Cencini, M.}, \au{Franchino, M.}, \au{Santamaria, F.} \& \au{Boffetta,
  G.}} \yr{2016}  \at{Centripetal focusing of gyrotactic phytoplankton}.
  \jt{J. Theor. Biol.}  \bvol{399},  \pg{62--70}.

\bibitem[Croze {\em et~al.\/}(2013)Croze, Sardina, Ahmed, Bees \&
  Brandt]{croze2013dispersion}
{\sc \au{Croze, O.~A.}, \au{Sardina, G.}, \au{Ahmed, M.}, \au{Bees, M.~A.} \&
  \au{Brandt, L.}} \yr{2013}  \at{Dispersion of swimming algae in laminar and
  turbulent channel flows: consequences for photobioreactors}.  \jt{J. R. Soc.
  Interface}  \bvol{10}~(81),  \pg{20121041}.

\bibitem[De~Lillo {\em et~al.\/}(2014)De~Lillo, Cencini, Durham, Barry,
  Stocker, Climent \& Boffetta]{delillo2014turbulent}
{\sc \au{De~Lillo, F.}, \au{Cencini, M.}, \au{Durham, W.M.}, \au{Barry, M.},
  \au{Stocker, R.}, \au{Climent, E.} \& \au{Boffetta, G.}} \yr{2014}
  \at{Turbulent fluid acceleration generates clusters of gyrotactic
  microorganisms}.  \jt{Phys. Rev. Lett.}  \bvol{112},  \pg{044502}.

\bibitem[Durham {\em et~al.\/}(2013)Durham, Climent, Barry, De~Lillo, Boffetta,
  Cencini \& Stocker]{durham2013turbulence}
{\sc \au{Durham, W.~M.}, \au{Climent, E.}, \au{Barry, M.}, \au{De~Lillo, F.},
  \au{Boffetta, G.}, \au{Cencini, M.} \& \au{Stocker, R.}} \yr{2013}
  \at{{Turbulence drives microscale patches of motile phytoplankton}}.
  \jt{Nat. Commun.}  \bvol{4},  \pg{2148}.

\bibitem[Durham {\em et~al.\/}(2009)Durham, Kessler \&
  Stocker]{durham2009disruption}
{\sc \au{Durham, W.~M.}, \au{Kessler, J.~O.} \& \au{Stocker, R.}} \yr{2009}
  \at{{Disruption of vertical motility by shear triggers formation of thin
  phytoplankton layers.}}  \jt{Science}  \bvol{323},  \pg{1067--70}.

\bibitem[Elgeti {\em et~al.\/}(2015)Elgeti, Winkler \&
  Gompper]{elgeti2015physics}
{\sc \au{Elgeti, J.}, \au{Winkler, R.~G.} \& \au{Gompper, G.}} \yr{2015}
  \at{Physics of microswimmers—single particle motion and collective
  behavior: a review}.  \jt{Rep. Progr. Phys.}  \bvol{78}~(5),  \pg{056601}.

\bibitem[Falkovich {\em et~al.\/}(2002)Falkovich, Fouxon \&
  Stepanov]{falkovich2002acceleration}
{\sc \au{Falkovich, G.}, \au{Fouxon, A.} \& \au{Stepanov, M.~G.}} \yr{2002}
  \at{Acceleration of rain initiation by cloud turbulence}.  \jt{Nature}
  \bvol{419}~(6903),  \pg{151}.

\bibitem[Fouxon \& Leshansky(2015)]{fouxon2015phytoplankton}
{\sc \au{Fouxon, I.} \& \au{Leshansky, A.}} \yr{2015}  \at{Phytoplankton's
  motion in turbulent ocean}.  \jt{Phys. Rev. E}  \bvol{92}~(1),  \pg{013017}.

\bibitem[Guasto {\em et~al.\/}(2010)Guasto, Johnson \&
  Gollub]{guasto2010oscillatory}
{\sc \au{Guasto, J.~S.}, \au{Johnson, K.~A.} \& \au{Gollub, J.~P.}} \yr{2010}
  \at{Oscillatory flows induced by microorganisms swimming in two dimensions}.
  \jt{Phys. Rev. Lett.}  \bvol{105}~(16),  \pg{168102}.

\bibitem[Guasto {\em et~al.\/}(2012)Guasto, Rusconi \&
  Stocker]{guasto2012fluid}
{\sc \au{Guasto, J.~S.}, \au{Rusconi, R.} \& \au{Stocker, R.}} \yr{2012}
  \at{Fluid mechanics of planktonic microorganisms}.  \jt{Annu. Rev. Fluid
  Mech.}  \bvol{44},  \pg{373--400}.

\bibitem[Gustavsson {\em et~al.\/}(2016)Gustavsson, Berglund, Jonsson \&
  Mehlig]{gustavsson2016preferential}
{\sc \au{Gustavsson, K.}, \au{Berglund, F.}, \au{Jonsson, P.R.} \& \au{Mehlig,
  B.}} \yr{2016}  \at{Preferential sampling and small-scale clustering of
  gyrotactic microswimmers in turbulence}.  \jt{Phys. Rev. Lett.}  \bvol{116},
  \pg{108104}.

\bibitem[Harvey {\em et~al.\/}(2015)Harvey, Menden-Deuer \&
  Rynearson]{harvey2015persistent}
{\sc \au{Harvey, E.~L.}, \au{Menden-Deuer, S.} \& \au{Rynearson, T.~A.}}
  \yr{2015}  \at{Persistent intra-specific variation in genetic and behavioral
  traits in the raphidophyte, heterosigma akashiwo}.  \jt{Front. Microbiol.}
  \bvol{6},  \pg{1277}.

\bibitem[Hill \& H{\"a}der(1997)]{hill1997biased}
{\sc \au{Hill, N.~A.} \& \au{H{\"a}der, D.~P.}} \yr{1997}  \at{A biased random
  walk model for the trajectories of swimming micro-organisms}.  \jt{J. Theor.
  Biol.}  \bvol{186}~(4),  \pg{503}.

\bibitem[Jeffery(1922)]{jeffery1922motion}
{\sc \au{Jeffery, G.~B.}} \yr{1922}  \at{The motion of ellipsoidal particles
  immersed in a viscous fluid}.  \jt{Proc. R. Soc. Lond. A}  \bvol{102}~(715),
  \pg{161--179}.

\bibitem[Kessler(1985)]{kessler1985hydrodynamic}
{\sc \au{Kessler, J.~O.}} \yr{1985}  \at{Hydrodynamic focusing of motile algal
  cells}.  \jt{Nature}  \bvol{313}~(5999),  \pg{218}.

\bibitem[Khurana {\em et~al.\/}(2011)Khurana, Blawzdziewicz \&
  Ouellette]{khurana2011reduced}
{\sc \au{Khurana, N.}, \au{Blawzdziewicz, J.} \& \au{Ouellette, N.~T.}}
  \yr{2011}  \at{Reduced transport of swimming particles in chaotic flow due to
  hydrodynamic trapping}.  \jt{Phys. Rev. Lett.}  \bvol{106}~(19),
  \pg{198104}.

\bibitem[Ki{\o}rboe(2008)]{kiorboe2008mechanistic}
{\sc \au{Ki{\o}rboe, T.}} \yr{2008} {\em A mechanistic approach to plankton
  ecology\/}.  \publ{Princeton University Press}.

\bibitem[Marchetti {\em et~al.\/}(2013)Marchetti, Joanny, Ramaswamy, Liverpool,
  Prost, Rao \& Simha]{marchetti2013hydrodynamics}
{\sc \au{Marchetti, M.~C.}, \au{Joanny, J.-F.}, \au{Ramaswamy, S.},
  \au{Liverpool, T.~B.}, \au{Prost, J.}, \au{Rao, M.} \& \au{Simha, R.~A.}}
  \yr{2013}  \at{Hydrodynamics of soft active matter}.  \jt{Rev. Mod. Phys.}
  \bvol{85},  \pg{1143}.

\bibitem[Nelson {\em et~al.\/}(2010)Nelson, Kaliakatsos \&
  Abbott]{nelson2010microrobots}
{\sc \au{Nelson, B.~J.}, \au{Kaliakatsos, I.~K.} \& \au{Abbott, J.~J.}}
  \yr{2010}  \at{Microrobots for minimally invasive medicine}.  \jt{Ann. Rev.
  Biomed. Eng.}  \bvol{12},  \pg{55--85}.

\bibitem[Pedley \& Kessler(1992)]{Pedley1992}
{\sc \au{Pedley, T.~J.} \& \au{Kessler, J.~O.}} \yr{1992}  \at{Hydrodynamic
  phenomena in suspensions of swimming microorganisms}.  \jt{Annu. Rev. Fluid
  Mech.}  \bvol{24},  \pg{313--358}.

\bibitem[Polin {\em et~al.\/}(2009)Polin, Tuval, Drescher, Gollub \&
  Goldstein]{Polin2009}
{\sc \au{Polin, M.}, \au{Tuval, I.}, \au{Drescher, K.}, \au{Gollub, J.~P.} \&
  \au{Goldstein, R.~E.}} \yr{2009}  \at{{Chlamydomonas swims with two "gears"
  in a eukaryotic version of run-and-tumble locomotion}}.  \jt{Science}
  \bvol{325},  \pg{487--90}.

\bibitem[Rusconi {\em et~al.\/}(2014)Rusconi, Guasto \&
  Stocker]{rusconi2014bacterial}
{\sc \au{Rusconi, R.}, \au{Guasto, J.~S.} \& \au{Stocker, R.}} \yr{2014}
  \at{Bacterial transport suppressed by fluid shear}.  \jt{Nat. Phys.}
  \bvol{10}~(3),  \pg{212}.

\bibitem[Santamaria {\em et~al.\/}(2014)Santamaria, De~Lillo, Cencini \&
  Boffetta]{Santamaria2014}
{\sc \au{Santamaria, F.}, \au{De~Lillo, F.}, \au{Cencini, M.} \& \au{Boffetta,
  G.}} \yr{2014}  \at{Gyrotactic trapping in laminar and turbulent kolmogorov
  flow}.  \jt{Phys. Fluids}  \bvol{26}~(11),  \pg{111901}.

\bibitem[Sengupta {\em et~al.\/}(2017)Sengupta, Carrara \&
  Stocker]{sengupta2017phytoplankton}
{\sc \au{Sengupta, A.}, \au{Carrara, F.} \& \au{Stocker, R.}} \yr{2017}
  \at{Phytoplankton can actively diversify their migration strategy in response
  to turbulent cues}.  \jt{Nature}  \bvol{543}~(7646),  \pg{555}.

\bibitem[Ten~Hagen {\em et~al.\/}(2014)Ten~Hagen, K{\"u}mmel, Wittkowski,
  Takagi, L{\"o}wen \& Bechinger]{ten2014gravitaxis}
{\sc \au{Ten~Hagen, B.}, \au{K{\"u}mmel, F.}, \au{Wittkowski, R.}, \au{Takagi,
  D.}, \au{L{\"o}wen, H.} \& \au{Bechinger, C.}} \yr{2014}  \at{Gravitaxis of
  asymmetric self-propelled colloidal particles}.  \jt{Nat. Commun.}  \bvol{5},
   \pg{4829}.

\bibitem[Thorpe(2007)]{Thorpe2007}
{\sc \au{Thorpe, S.~A.}} \yr{2007} {\em An introduction to ocean turbulence\/}.
   \publ{Cambridge University Press}.

\bibitem[Torney \& Neufeld(2007)]{Torney2007}
{\sc \au{Torney, C.} \& \au{Neufeld, Z.}} \yr{2007}  \at{{Transport and
  Aggregation of Self-Propelled Particles in Fluid Flows}}.  \jt{Phys. Rev.
  Lett.}  \bvol{99},  \pg{078101}.

\bibitem[Torney \& Neufeld(2008)]{Torney2008}
{\sc \au{Torney, C.} \& \au{Neufeld, Z.}} \yr{2008}  \at{Phototactic clustering
  of swimming microorganisms in a turbulent velocity field}.  \jt{Phys. Rev.
  Lett.}  \bvol{101},  \pg{078105}.

\bibitem[Zhan {\em et~al.\/}(2014)Zhan, Sardina, Lushi \&
  Brandt]{zhan2014accumulation}
{\sc \au{Zhan, C.}, \au{Sardina, G.}, \au{Lushi, E.} \& \au{Brandt, L.}}
  \yr{2014}  \at{Accumulation of motile elongated micro-organisms in
  turbulence}.  \jt{J. Fluid Mech.}  \bvol{739},  \pg{22--36}.

\end{thebibliography}

\end{document}